\documentclass[10pt,amsmath,amssymb,nofootinbib,twoside,twocolumn,superscriptaddress,floats,floatfix,aps,prl,preprintnumbers]{revtex4-2}
\usepackage[british]{babel}
\usepackage{xcolor}
\usepackage{booktabs}
\usepackage{latexsym}
\usepackage{amssymb}
\usepackage{amsmath}
\usepackage{graphicx}
\usepackage{tabularx}
\usepackage{cancel}
\usepackage{hyperref}
\usepackage{aas_macros}
\hypersetup{
    colorlinks=true,
    linkcolor=black,
    citecolor=blue,
    filecolor=black,
    urlcolor=blue,
    pdfauthor={M. Cadoni, M. De Laurentis, R. Della Monica, I. De Martino, M. Oi, and A. P. Sanna},
    pdftitle={Testing a novel nonsingular black-hole model with S2 star}
}
\usepackage[capitalise]{cleveref}
\usepackage[utf8]{inputenc}
\usepackage[normalem]{ulem}
\def\beq#1\eeq{\begin{align}#1\end{align}}

\newcommand{\dd}{\text{d}}

\newcommand{\sce}{\mathit{e}}

\newcommand{\rH}{r_\text{H}}
\newcommand{\Th}{T_\text{H}}
\newcommand{\barl}{\bar{\ell}}
\newcommand{\ellC}{\ell_\text{c}}
\newcommand{\Ord}{\mathcal{O}}
\newcommand{\rext}{r_\text{H,ext}}
\newcommand{\dotx}{\dot{x}}
\newcommand{\curlyE}{\mathcal{E}}

\begin{document}

\newcommand{\UniCa}{\affiliation{Dipartimento di Fisica, Universit\`a di Cagliari, Cittadella Universitaria, 09042 Monserrato, Italy}}
\newcommand{\INFNCa}{\affiliation{INFN, Sezione di Cagliari, Cittadella Universitaria, 09042 Monserrato, Italy}}
\newcommand{\UniNa}{\affiliation{Dipartimento di Fisica ``E. Pancini'', Universit\`a di Napoli ``Federico II'', Compl. Univ. di Monte S. Angelo, Edificio G, via Cintia, I-80126, Napoli, Italy}}
\newcommand{\INFNNa}{\affiliation{INFN, Sezione di Napoli, Compl. Univ. di Monte S. Angelo, Edificio G, via Cintia, I-80126, Napoli, Italy}}
\newcommand{\Salamanca}{\affiliation{Departamento de F\'isica Fundamental, Universidad de Salamanca, Plaza de la Merced, s/n, E-37008 Salamanca, Spain}}

\author{M.~Cadoni}
\email{mariano.cadoni@ca.infn.it}
\UniCa\INFNCa

\author{M.~De~Laurentis}
\email{felicia@na.infn.it}
\UniNa\INFNNa

\author{I.~De~Martino}
\email{ivan.demartino@usal.es}
\Salamanca

\author{R.~Della~Monica}
\email{rdellamonica@usal.es}
\Salamanca

\author{M.~Oi}
\email{mauro.oi@ca.infn.it}
\UniCa\INFNCa

\author{A.~P.~Sanna}
\email{asanna@dsf.unica.it}
\UniCa\INFNCa

\title{Are nonsingular black holes with super-Planckian hair ruled out by S2 star data?}

\begin{abstract}
    We propose a novel nonsingular black-hole spacetime representing a strong deformation of the Schwarzschild solution with mass $M$ by an additional hair $\ell$, which may be hierarchically larger than the  Planck scale. Our black-hole model presents a de Sitter core and $\mathcal{O}(\ell^2/r^2)$ slow-decaying corrections to the Schwarzschild solution. Our black-hole solutions are thermodynamically preferred when $0.2 \lesssim \ell/GM \lesssim \, 0.3$ and are characterized by strong deviations in the orbits of test particles from the Schwarzschild case. In particular, we find corrections to the perihelion precession angle scaling linearly with $\ell$. We test our model using the available data for the orbits of the S2 star around $\text{SgrA}^*$. These data strongly constrain the value of the hair $\ell$, casting an upper bound on it of $\sim \, 0.47 \, GM$, but do not rule out the possible existence of regular black holes with super-Planckian hair. 
\end{abstract}

\maketitle

\emph{Introduction}.---Since its formulation, General Relativity (GR) has been widely tested in several different contexts. In particular, one of its most intriguing predictions is the existence of black holes, whose possible presence has been tested both directly \cite{EventHorizonTelescope:2019dse,EventHorizonTelescope:2022wkp,LIGOScientific:2016aoc} and indirectly \cite{Eckart:1997em,Ghez:2000ay}. Although black-hole imaging, gravitational wave and iron-line observations are compatible with the presence of Kerr black holes \cite{Bambi:2011mj,Tripathi:2021rqs,EventHorizonTelescope:2019dse,EventHorizonTelescope:2022xqj,LIGOScientific:2016aoc}, there is still room for small deviations, which could be tested with present and future experiments. For this reason, there has been an increasing interest in studying black-hole mimickers. These objects share some properties with GR solutions but allow for a different phenomenology at the horizon scale, which, if observed, would represent a smoking gun for deviations from GR.

Regular black holes are among the most fascinating mimickers. Contrary to classical black holes which present a singularity at their core \cite{PhysRevLett.14.57,Hawking:1970zqf}, indicating the breakdown of the classical theory, these objects are completely regular everywhere. This point is particularly significant, since we expect a quantum theory of gravity to resolve the classical singularity problem. Although there have been some attempts to capture the main properties of the fundamental theory \cite{Mathur:2005zp,Perez:2017cmj,Buoninfante:2019swn}, a clear understanding of its dynamics and of the mechanism leading to the formation of these regular spacetimes is still lacking. Consequently, mainly bottom-up approaches have been followed until now \cite{bardeen1968proceedings,Hayward:2005gi,Cadoni:2022chn,Franzin:2022wai,Mazza:2021rgq,Simpson:2019mud,Simpson:2021dyo,Sebastiani:2022wbz,Jusufi:2022cfw}, in which one usually modifies GR solutions to test possible deviations in a phenomenological fashion. This is particularly suitable, for instance, to study orbits of test particles \cite{DeMartino2021,DellaMonica2022b}, and it has been used to investigate modifications to the black-hole shadow \cite{Younsi:2016azx,Konoplya:2016jvv,Johannsen2011,PsaltisEHT2020,EventHorizonTelescope:2022xqj,Ghosh:2022gka}. An intriguing feature of nonsingular black-hole models is the presence of a new length scale (hair) $\ell$, which can be hierarchically larger than the Planck scale \cite{Cadoni:2022chn}. An important question to be answered is whether such models with super-Planckian hair can be excluded by present experimental data. 

In this work, following this type of approach, we build a novel regular black-hole metric, belonging to the general class of models explored in Ref.~\cite{Cadoni:2022chn}. The corrections to the Schwarzschild spacetime decrease sufficiently slow to be experimentally observable also at great distances from the horizon. Therefore, we test our model with the orbital motion of the S2 star around the compact radio source Sagittarius A* (SgrA* \cite{EventHorizonTelescope:2022wkp}) in the Galactic Center (GC) \cite{Schodel2002, Ghez2003}. 

\emph{The model}.---We start from a static and spherically symmetric spacetime written in Schwarzschild coordinates $(t,r,\theta,\phi)$ \footnote{Throughout the paper, we shall use natural units in which $c=\hbar=k_\text{B}=1$, unless otherwise specified.} 
\begin{subequations}
\beq
\dd s^2 & = - f(r) \dd t^2 +  \frac{\dd r^2}{f(r)} + r^2 \dd\Omega^2\, ,\label{eq:general_metric}\\
f(r) & = 1-\frac{2G m(r)}{r}, \quad m(r) = 4\pi \int_0^r \dd\tilde r \, \tilde r^2 \, \rho(\tilde r)\, , \label{eq:density}
\eeq
\end{subequations}
where $\dd\Omega^2 = \dd \theta^2 + \sin^2\theta\,\dd\phi^2$ is the $2\text{-sphere}$ line element, while $m(r)$ is the Misner-Sharp (MS) mass, given by a density profile $\rho(r)$. Some profiles  of the MS mass \eqref{eq:density} give a general class of solutions of GR sourced by a anisotropic fluid, with  equation of state $p_{r}=-\rho$, where $p_r$ is the radial component of the fluid pressure \cite{Cadoni:2022chn}. Due to Birkhoff's theorem, the only vacuum solution of GR of the form \eqref{eq:general_metric} is the Schwarzschild one, for which $m(r) = M$, being $M$ the Arnowitt-Deser-Misner mass of the object. This solution is plagued by a curvature singularity at $r=0$. A widely used approach to eliminate the latter is imposing a de Sitter (dS)-like behavior of the metric in the core, which translates to require $f(r) = 1 - \alpha r^2/\ell^2 + \Ord(r^3/\ell^3)$ for $r \ll GM$, where $\alpha$ is some positive dimensionless constant determining the inner dS length. Here, we introduced an additional length scale $\ell$, which is responsible for the smearing of the classical singularity and can be interpreted as an additional hair of the black hole. Moreover, asymptotic flatness at infinity requires $f(r) = 1 - 2GM / r + \beta \ell^2 / r^2 + \Ord(\ell^3/r^3)$ for $r \gg GM$, where $\beta$ is, again, a dimensionless constant. These two minimal requirements characterize a wide class of regular models, whose general properties were investigated in Ref.~\cite{Cadoni:2022chn}.

In this paper we select the particularly simple case belonging to such general class
\beq\label{DMinspiredmetric}
f(r) = 1- \frac{2GM r^2}{(r + \ell)^3}\, ,
\eeq
for which $\alpha = 2GM/\ell$ and $\beta = 6GM/\ell$. The Schwarzschild spacetime is recovered in the limit $\ell \to 0$. 

The main reason for selecting the metric \eqref{DMinspiredmetric} is that it gives $\mathcal{O}(\ell^2/r^2)$ corrections to the Schwarzschild geometry, which are the most powerful deformations still compatible with Schwarzschild asymptotics. We therefore expect strong deviations from the standard GR phenomenology, having a clear and potentially observable experimental signature at least when the hair $\ell$ is super-Planckian. Until now, only models that have at most order $\mathcal{O}(\ell^4/r^4)$ (like, e.g., the Hayward black hole \cite{Hayward:2005gi}) or exponentially-suppressed corrections \cite{Nicolini:2005vd} have been investigated (see also Ref.~\cite{Cadoni:2022chn} and references therein). The metric has also a nice astrophysical analogy. In fact, the density associated with the MS mass in \cref{DMinspiredmetric} reads $4\pi\rho(r) = 3M\ell/(r + \ell)^4$, whose large $r$ behavior, $\sim r^{-4}$, reproduces  that of some density profiles of dark matter in elliptical and spherical galaxies, like the Hernquist \cite{Hernquist:1990be} or Jaffe \cite{Jaffe:1983iv} ones.

The spacetime structure is the same of the general class of models discussed in Ref. \cite{Cadoni:2022chn}. It has an outer event and an inner Cauchy horizons, respectively located at $r_+$ and $r_-$, for which $f(r_\pm)=0$. The two horizons coincide at $\rext = 2\ellC$ for the critical value $\ell=\ellC \equiv 8GM/27$, and the black hole becomes extremal. The two horizons remain separate for $\ell < \ellC$, while they disappear for $\ell>\ellC$, which corresponds to a compact horizonless object.

\emph{Black-hole thermodynamics}.---The mass $M$ and the Hawking temperature $\Th$ of the black-hole model, as functions of the outer horizon $r_+\equiv\rH$ and $\ell$, read
\begin{subequations}\label{MTrH}
    \beq
        \label{MrH}
        M = \frac{(\ell+\rH)^3}{2 G \rH^2},\quad
        \Th = \frac{\rH-2 \ell}{4 \pi  \rH (\ell+\rH)}\, .
    \eeq
\end{subequations}

As it is typical, the temperature is zero at extremality. Unlike the Schwarzschild case, the temperature is nonmonotonic and has a maximum located at $r_\text{H,peak} = \left(2 + \sqrt{6} \right) \, \ell$. This implies the presence of metastable states, i.e., configurations with the same temperature, but different horizon radii and, therefore, thermodynamic properties. This can be seen by computing the specific heat
\beq\label{CH}
    C_\text{H} = \frac{\dd M}{\dd\Th}=-\frac{2 \pi  (\rH-2\ell) (\ell+\rH)^4}{G \rH \left(-2 \ell^2-4 \ell \rH+\rH^2\right)}\, .
\eeq
It diverges at $\rH = r_\text{H,peak}$, signaling the onset of a second-order phase transition, separating the models into two branches. For $\rH > r_\text{H,peak}$, i.e., $\ell/GM \leq \frac{4}{9}(3-\sqrt{6}) \sim 0.245$, $C_\text{H} <0$: this is the branch of thermodynamically unstable  holes. For $\rext < \rH < r_\text{H,peak}$, i.e., $0.245\, GM \le \ell \le \ellC$, $C_\text{H}\geq 0$, we have instead the branch of stable black holes. This branch includes also the extremal configuration, for which $C_\text{H} = 0$. In order to assess which branch is thermodynamically preferred, we computed the free energy $\mathcal{F} = M - \Th S$, using both \cref{MrH} and the entropy formula in terms of the integral of the mass function, proposed in Ref.~\cite{Cadoni:2022chn}. We found that the thermodynamically stable black holes, with $\ell \sim \ellC$, attain the least free energy and are, thus, thermodynamically preferred with respect to their unstable counterparts with $\ell \ll GM$. 

\textit{Orbits of test particles}.---The orbits of test particles (geodesics) in the spacetime \eqref{DMinspiredmetric} can be analyzed using a standard procedure. One writes the Lagrangian $\mathcal{L}=g_{\mu\nu}\dotx^\mu\dotx^\nu$ (the dot means differentiation with respect to an affine parameter $\lambda$) and uses the isometries of the metric and the associated conserved quantities (energy $\curlyE$ and angular momentum). Without loss of generality, we consider, in our analysis, orbits in the equatorial plane, for which we have
\beq\label{eq:rdot_geodesics}
    \dot{r}^2 + f(r)\left(\epsilon^2 + \frac{L^2}{r^2}\right) = \curlyE^2.
\eeq
$\epsilon$ is equal to $\pm 1$ or $0$ for timelike and null orbits respectively, and $L$ denotes the angular momentum about an axis normal to the invariant plane. The qualitative behavior of both null and timelike trajectories can be studied by analyzing the effective potential $V(r) = f(r)\left(\epsilon^2 + L^2/r^2\right)$. For null trajectories, $V(r)$ has a minimum in the region $r_- \leq r \leq r_+$ and a maximum located at $r>r_+$, the latter corresponding to an unstable circular orbit (light ring). The light ring radius, equal to $3\, GM$ for the Schwarzschild spacetime, decreases linearly with $\ell$ for small values of the latter, going rapidly to zero at $\ell\sim 0.317 \, GM$. A similar analysis shows that, for timelike trajectories, the potential has up to three extrema, depending on the values of $\ell$ and $L$. 
A first minimum is located between the outer and the inner horizon. The other two extrema represent a maximum and a minimum, which correspond  to the marginally bound and the stable circular orbits, respectively.

Starting from \cref{eq:rdot_geodesics}, we can now write an equation for the orbits in the variable $u = 1/r$
\beq\label{eq:orbits}
    {u'}^2 + f(u)\left(\frac{\epsilon^2}{L^2} + u^2\right) = \frac{\curlyE^2}{L^2},
\eeq
where a prime indicates differentiation with respect to $\phi$, and $f(u) = 1 - 2GMu/(1 + u\ell)^3$.  We introduce the dimensionless quantities $\sigma = (GM/L)^2$, $\barl = \ell/GM$ and the new variable $\xi =  GMu/\sigma$. When the parameter $\sigma$ is small (e.g., $\sigma = 1.88\times10^{-4}$ for the S2 star), we can solve \cref{eq:orbits} perturbatively looking for a solution of the form $\xi = \xi_0 + \sigma\xi_1 + \Ord(\sigma^2)$.

The zeroth order gives the Newtonian orbits, while at first order we have  the equation
\beq
\xi_1'' + \xi_1 = 3\xi_0(\xi_0 - 2\barl)\, .
\eeq
Neglecting subdominant contributions, and defining $\gamma \equiv 1 - \barl$, we see that the full solution becomes $\xi \simeq 1 + \sce\cos\left[\left(1 - 3\gamma\sigma\right)\phi\right]$, where $\sce$ is the orbit eccentricity. The orbit precedes of an angle
\beq\label{eq:precession}
\Delta\phi \simeq 6\pi\sigma\gamma\, .
\eeq
\begin{figure}[t]
    \centering
    \includegraphics[width=\columnwidth]{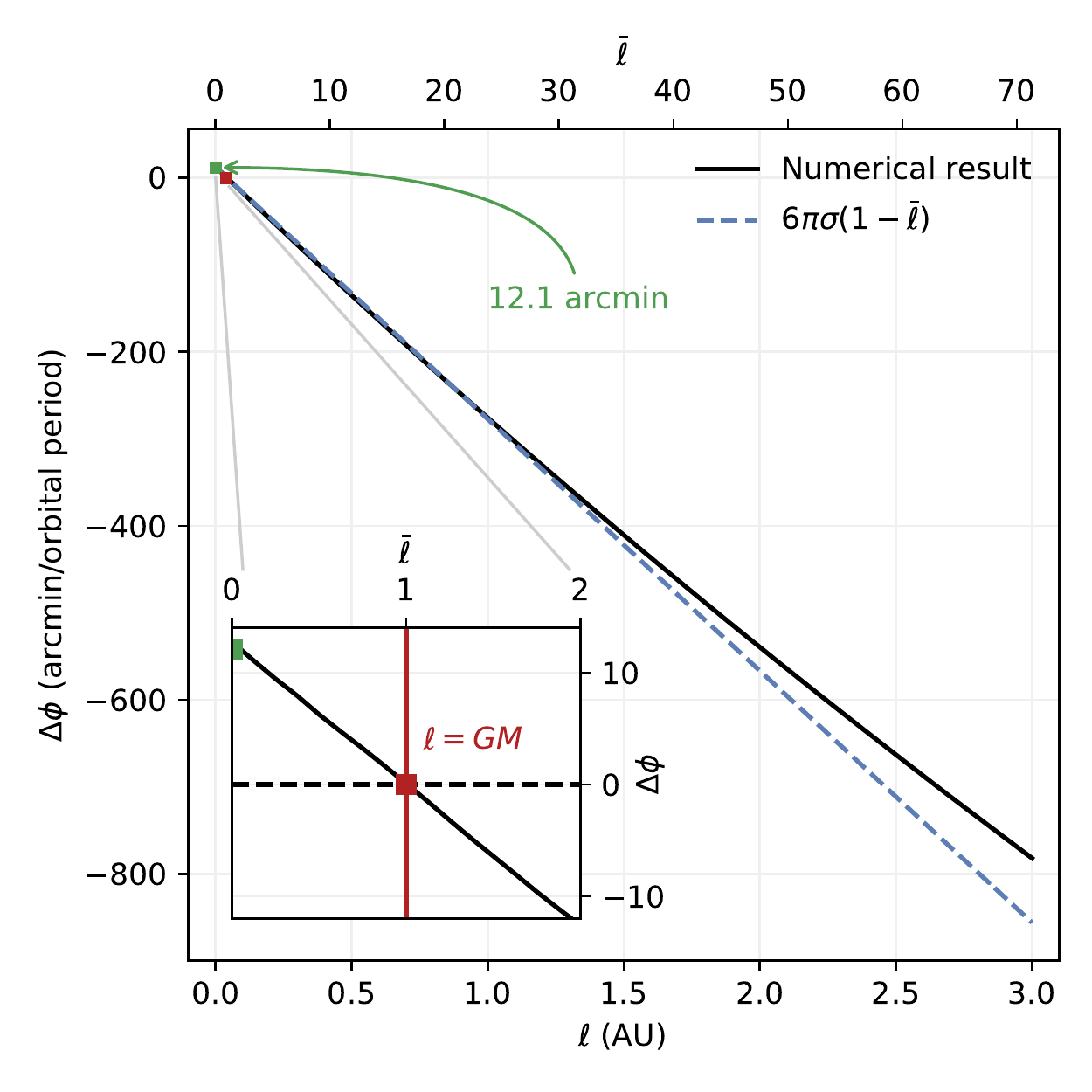}
    \caption{Orbital precession for the S2 star for our model as a function of $\ell$. The black solid line reports the results of our numerical integration of the geodesic equations in \cref{eq:orbits}, the blue dashed line represents our first-order perturbative prediction in \cref{eq:precession}. For $\ell = 0$ we obtain the Schwarzschild precession of 12.1 arcmin per orbital period observed in \cite{Gravity2020}. Moreover, we are able to confirm the perturbative result for small values of $\ell$ up to a few tens of gravitational radii, where departure from the linear trend in \cref{eq:precession} is exhibited by the numerical prediction. Finally, we confirm numerically that for $\ell\geq GM$, the orbital precession becomes retrograde. }
    \label{fig:precession}
\end{figure}
While the usual Schwarzschild result is obtained for $\barl \to 0$, we see that $\Delta\phi$ decreases linearly with the additional hair and it becomes retrograde for $\barl > 1$. This is an interesting feature, not present in the orbits of the Schwarzschild spacetime, that can be used  to strongly constrain the model.

In order to probe the possible existence of the hair $\ell$ in \cref{DMinspiredmetric}, we have developed an orbital model for the S2 star in the GC based on the numerical integration of \cref{eq:orbits}. In particular, one can recast the energy $\mathcal{E}$ and the angular momentum $L$ in \cref{eq:orbits} in terms of the classical Keplerian elements: the semi-major axis $a$, the eccentricity $e$, the time of pericenter passage $t_p$ and the orbital period $T$ (which can, in fact, be derived from $M$ and $a$ through Kepler's third law). A choice of these parameters uniquely identifies a Keplerian ellipse on the equatorial plane, that we assume to osculate the real trajectory of the star at a given time. We hence make use of such ellipse to set the initial conditions at a time $t_0$ that, without loss of generality, we fix to be the last time of apocentre passage for S2 in $\sim$2010.35.

Starting from such initial conditions, we integrate the geodesic equations numerically by means of a fourth-order Runge-Kutta scheme, over approximately two orbital periods, covering the time range that spans from 1990 to 2017. 
Finally, to compare our synthetic orbit with public data, we need to reconstruct the observable quantities for S2, i.e., the astrometric sky-projected position over time for an Earth-based observatory and the spectroscopically-measured line-of-sight velocity of the star. To this aim, we perform a geometric projection of the star's trajectory in the observer reference frame by means of the Thiele-Innes formulas computed from the three angular orbital elements: $i$, the orbital inclination; $\Omega$, the longitude of the ascending node; $\omega$, the argument of the pericenter. Additionally, for the spectroscopic observables, we take into account the post-Newtonian time-dilation effects on the light emitted by the star, namely the special-relativistic transverse and longitudinal Doppler effect and the general-relativistic gravitational redshift (for more details on how such quantities can be appropriately accounted for, we refer to  previous works on the subject \cite{DellaMonica2022a, DellaMonica2022b}). 

The orbital precession is naturally taken into account on our synthetic orbits since we directly integrate the fully-relativistic geodesic equations derived from \cref{eq:general_metric,DMinspiredmetric}. As a matter of fact, our numerically-integrated orbit allows us to effectively validate the perturbative results in \cref{eq:precession} by computing the orbital precession $\Delta\phi$ as the angle spanned by the star between two subsequent radial turning points.
In \cref{fig:precession} we report a comparison between the numerically computed orbital precession (black solid line) for the S2 star as a function of the parameter $\ell$, once all the other Keplerian elements have been fixed to the ones of S2, as derived from the analysis in \cite{Gillessen2017} (based on a Newtonian orbital model). For $\ell = 0$ we obtain the Schwarzschild precession of 12.1 arcmin per orbital period observed in \cite{Gravity2020}. The linearly decreasing trend predicted by our perturbative analysis (dashed blue line) is confirmed up to a few tens of gravitational radii for the parameter $\ell$, where our numerical predictions start to depart. Moreover, we are able to confirm numerically the prediction that the orbital precession becomes retrograde for $\ell\geq GM$.

\emph{Constraining the model with S2 orbital data}.---%
\begin{figure}[t]
    \centering
    \includegraphics[width=\columnwidth]{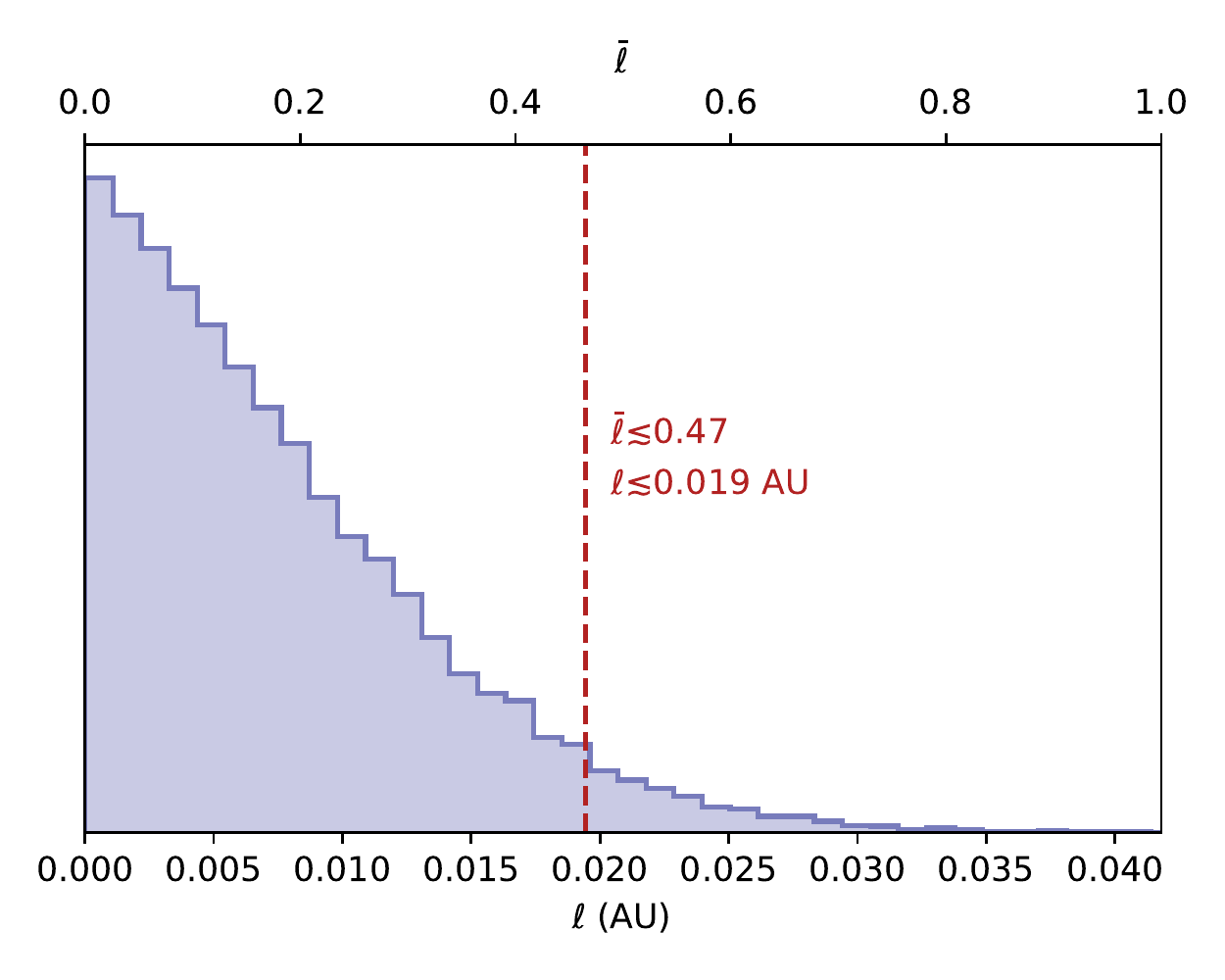}
    \caption{Marginalized posterior probability distribution for the parameter $\ell$ resulting from our MCMC analysis. We are able to constrain the parameter $\ell$ by providing a 95\% confidence level upper limit of $\ell\lesssim0.019$ AU, corresponding to $\barl \lesssim 0.47$. }
    \label{fig:posterior}
\end{figure}%
We have explored the parameter space of our orbital model for the S2 star, using the publicly available orbital data for S2. In particular, we have used near-infrared astrometric positions and radial velocities, coming from 25 years of uninterrupted monitoring of stellar orbits in the GC between $\sim1992$ and $\sim2017$, presented in \cite{Gillessen2017}.
These data do not cover the last S2 pericenter passage in May 2018 nor its subsequent motion observed by the GRAVITY Collaboration \cite{Gravity2020, Gravity2022}. Information provided by such portion of the orbit is crucial in the deed of constraining the gravitational field of SgrA* \cite{Grould2017}. However, as demonstrated in previous works \cite{DeMartino2021, DellaMonica2022a}, one can consider orbital data that do not cover the pericenter passage and, then, add as a single datapoint the precession measurement ($f_{\rm SP} = 1.10\pm0.19$ from \cite{Gravity2020}, where $f_{\rm SP}=0$ corresponds to a non-preceeding ellipse from Newtonian gravity and $f_{\rm SP}=1$ corresponds to the GR rate of orbital precession for the Schwarzschild spacetime).

Besides the knowledge of $\ell$, the full orbital model requires the knowledge of all the seven Keplerian parameters 
(which implies leaving the mass of the central object as a free parameter, as well), along with the observer galactocentric distance $D$ and 5 reference frame parameters \cite{Gillessen2017, Plewa2015}. The 14-dimensional parameter space has been explored through a Markov-Chain Monte Carlo (MCMC) algorithm implemented in \cite{ForemanMackey2013}. For the sake of generality, we have employed uniform priors on all the Keplerian parameters of our orbital model corresponding to an interval centered on the best-fit values from \cite{Gillessen2017}, with amplitude being 10 times the corresponding observational uncertainty. For the reference frame parameters, on the other hand, we have taken priors from the independent analysis in \cite{Plewa2015}. The interval for the hair $\ell$ has been set heuristically between 0 and 5 AU, corresponding to over 100 gravitational radii of the central source. The likelihood adopted for our analysis is the following
\begin{align}
	\log \mathcal{L} =& -\frac{1}{2}\sum_i\biggl(\frac{\textrm{R.A.}_i-\textrm{R.A.}_{{\rm obs}, i}}{\sqrt{2}\sigma_{{\rm R.A.},i}}\biggr)^2+\nonumber\\&-\frac{1}{2}\sum_i\biggl(\frac{\textrm{Dec}_i-\textrm{Dec}_{{\rm obs}, i}}{\sqrt{2}\sigma_{{\rm Dec},i}}\biggr)^2+\nonumber\\
	& -\frac{1}{2}\sum_i\biggl(\frac{\textrm{RV}_i-\textrm{RV}_{{\rm obs}, i}}{\sqrt{2}\sigma_{{\rm RV},i}}\biggr)^2+\nonumber\\&-\frac{1}{2}\biggl(\frac{\Delta\phi/\Delta\phi_{\rm GR}-f_{{\rm SP}}}{\sqrt{2}\sigma_{f_{\rm SP}}}\biggr)^2,
	\label{eq:likelihood}
\end{align}
where R.A., Dec and RV correspond to the sky-projected right ascension and declination of S2 and its radial velocity, respectively, while $\Delta \phi_\text{GR}$ is the precession angle predicted for the Schwarzschild spacetime. The subscript \emph{obs} represents the observed quantity at the $i\text{-th}$ epoch, and the $\sigma$'s are the corresponding observational uncertainties. As done in \cite{DeMartino2021}, the factors $\sqrt{2}$ in the denominators are introduced in order not to double count data points when considering the last term with the orbital precession (that has been derived with the same dataset).

The results of our posterior analysis are presented in \cref{tab:posterior} where the medians and the 68\% confidence level intervals for each bounded parameter are reported. They agree within 1$\sigma$ with previous results in the literature \cite{Gillessen2017}.  We are able to place an upper limit $\ell \lesssim 0.019$ AU (corresponding to $\ell\lesssim 0.47 \, GM$) at 95\% confidence level on the additional hair, whose marginalized posterior distribution is shown in \cref{fig:posterior}. Finally, we tested deviations from a Schwarzschild black hole using a mock catalogue (for more details see \cite{DellaMonica2022a}) that mirrors future GRAVITY observations of S2, and we proved the ability of GRAVITY to improve the upper limit on the hair $\ell$ derived in this work by a factor $\sim 10$.
\begin{table*}[ht]
    \centering
    \setlength{\tabcolsep}{20pt}
    \renewcommand{\arraystretch}{1.5}
    \begin{tabular}{|lc|lc|}
        \hline
        \textbf{Parameter (units)} & \textbf{Best-fit} & \textbf{Parameter (units)} & \textbf{Best-fit} \\ \hline
        $D$ (kpc) & $8.24\pm0.22$ & $\omega$ ($^\circ$) & $65.23_{-0.77}^{+0.78}$ \\
        $T$ (yr) & $16.050\pm0.028$ & $x_0$ (mas) & $0.26\pm0.16$ \\
        $t_p$ (yr) & $2018.379\pm0.024$ & $y_0$ (mas) & $-0.04_{-0.20}^{+0.19}$ \\
        $a$ (as) & $0.1249_{-0.0010}^{+0.0011}$ & $v_{x,0}$ (mas/yr) & $0.071_{-0.052}^{+0.053}$ \\
        $e$ & $0.8828\pm0.0024$ & $v_{z,0}$ (mas/yr) & $0.092\pm0.062$ \\
        $i$ ($^\circ$) & $134.42_{-0.49}^{+0.48}$ & $v_{z,0}$ (km/s) & $-3.4\pm4.5$ \\
        $\Omega$ ($^\circ$) & $226.75_{-0.82}^{+0.83}$ & $\barl$ & $\lesssim 0.47$ {\footnotesize (95\% c.l.)} \\ \hline
    \end{tabular}
    \caption{Results of our posterior analysis for the 14 parameters of our orbital model for the S2 star. In particular, for the bounded parameters, we derived and reported the 68\% confidence interval around the median of the marginalized distributions. On the other hand, for the parameter $\ell$, our analysis yields an upper limit which, at 95\% confidence level, is given by $\ell \lesssim 0.019$ AU (corresponding to the dimensionless value $\barl\lesssim 0.47$).}
    \label{tab:posterior}
\end{table*}

\emph{Discussion and conclusions}.---We proposed a novel regular black-hole spacetime characterized by an additional hair $\ell$, responsible for the smearing of the classical singularity. Our metric is designed to have the strongest allowed corrections at infinity, with respect to the standard Schwarzschild solution. Due to this property, we have been able to test such geometry with the orbits of the S2 star around SgrA$^\ast$ in the GC. We have found that corrections to the perihelion precession angle scale linearly with $\ell$. We have also been able to constrain the parameter $\ell$ with a MCMC analysis, whose result is the upper bound $\ell \lesssim 0.47 \, GM$. This rules out most horizonless solutions but allows the existence of thermodynamically stable regular black holes, i.e., models with $0.254 \leq \ell/GM \leq 0.296$. Our results show that regular black holes with super-Planckian hair are not excluded by the S2 star observational data. The actual proof of the existence of our nonsingular black holes requires the measurement of $\ell$-dependent deviations from GR, which could be detected by observations at the light-ring scale, e.g., with the black hole shadow or gravitational wave experiments. Nonetheless, we expect our model not to be a good approximation at that scale, and that a rotating generalization could be necessary.

\emph{Addendum}.---After this manuscript was completed, we were informed that the metric \eqref{DMinspiredmetric} was first proposed in Ref.~\cite{Lan:2022qbb} and used to investigate the asymptotic quasinormal modes of regular black holes.

\emph{Acknowledgements}.---We thank S. Zerbini for pointing our attention to Ref.~\cite{Lan:2022qbb}. RDM acknowledges support from Consejeria de Educación de la Junta de Castilla y León and from the Fondo Social Europeo. IDM acknowledges support from Ayuda IJCI2018-036198-I funded by MCIN/AEI/10.13039/501100011033 and: FSE “El FSE invierte en tu futuro” o “Financiado por la Unión Europea “NextGenerationEU”/PRTR. IDM is also supported by the project  PID2021-122938NB-I00 funded by the Spanish "Ministerio de Ciencia e Innovación" and FEDER “A way of making Europe", and by the project SA096P20 Junta de Castilla y León. MDL acknowledges the support of Istituto Nazionale di Fisica Nucleare (INFN) {\it iniziative specifiche} QGSKY and TEONGRAV.

\bibliography{refs}
\end{document}